\documentclass[journal]{IEEEtran}
\usepackage{float}
\usepackage{cite}										
\usepackage{amsmath}									
\usepackage{mathtools}
\usepackage{graphicx}

\usepackage{epstopdf}
\usepackage{float}
\usepackage{bm,upgreek}
\usepackage{amsfonts}
\usepackage{enumitem}
\usepackage{pgfplots}
\pgfplotsset{compat=1.5}
\usepackage{pgfplotstable}
\usepgfplotslibrary{statistics}

\emergencystretch=\maxdimen
\allowdisplaybreaks
\ifCLASSINFOpdf
\else
\fi

\ifCLASSOPTIONcompsoc
  \usepackage[caption=false,font=normalsize,labelfont=sf,textfont=sf]{subfig}
\else
 \usepackage[caption=false,font=footnotesize]{subfig}
\fi

\makeatletter
\pgfplotsset{
    boxplot prepared from table/.code={
        \def\tikz@plot@handler{\pgfplotsplothandlerboxplotprepared}%
        \pgfplotsset{
            /pgfplots/boxplot prepared from table/.cd,
            #1,
        }
    },
    /pgfplots/boxplot prepared from table/.cd,
        table/.code={\pgfplotstablecopy{#1}\to\boxplot@datatable},
        row/.initial=0,
        make style readable from table/.style={
            #1/.code={
                \pgfplotstablegetelem{\pgfkeysvalueof{/pgfplots/boxplot prepared from table/row}}{##1}\of\boxplot@datatable
                \pgfplotsset{boxplot/#1/.expand once={\pgfplotsretval}}
            }
        },
        make style readable from table=lower whisker,
        make style readable from table=upper whisker,
        make style readable from table=lower quartile,
        make style readable from table=upper quartile,
        make style readable from table=median,
        make style readable from table=lower notch,
        make style readable from table=upper notch
}
\makeatother

\pgfplotscreateplotcyclelist{list2}{%
{color=red}}
\pgfplotsset{
  grid style = {
    line cap = round,
    gray!20,
    line width = 0.1pt
  }
}

\pgfplotsset{legend image with text/.style={legend image code/.code={%
\node[anchor=west, align=right] at (0.0cm,0cm) {#1};}},}

\begin{document}
\title{Linear State Estimation via 5G C-RAN Cellular Networks using Gaussian Belief Propagation}

\author{Mirsad~Cosovic,
        Dejan Vukobratovic, Vladimir Stankovic% <-this % stops a space

\thanks{M. Cosovic is with Schneider Electric DMS NS, Novi Sad, Serbia (e-mail: mirsad.cosovic@schneider-electric-dms.com). D. Vukobratovic is with Department of Power, Electronic and Communications Engineering, University of Novi Sad, Novi Sad, Serbia (e-mail: dejanv@uns.ac.rs). V. Stankovic is with Department of Electronic and Electrical Engineering, University of Strathclyde, Glasgow, UK (e-mail:vladimir.stankovic@eee.strath.ac.uk).}}

\maketitle

\begin{abstract}
Machine-type communications and large-scale information processing architectures are among key (r)evolutionary enhancements of emerging fifth-generation (5G) mobile cellular networks. Massive data acquisition and processing will make 5G network an ideal platform for large-scale system monitoring and control with applications in future smart transportation, connected industry, power grids, etc. In this work, we investigate a capability of such a 5G network architecture to provide the state estimate of an underlying linear system from the input obtained via large-scale deployment of measurement devices. Assuming that the measurements are communicated via densely deployed cloud radio access network (C-RAN), we formulate and solve the problem of estimating the system state from the set of signals collected at C-RAN base stations. Our solution, based on the Gaussian Belief-Propagation (GBP) framework, allows for large-scale and distributed deployment within the emerging 5G information processing architectures. The presented numerical study demonstrates the accuracy, convergence behavior and scalability of the proposed GBP-based solution to the large-scale state estimation problem.
\end{abstract}

%\begin{IEEEkeywords}
%\end{IEEEkeywords}

\IEEEpeerreviewmaketitle

\section{Introduction}

With transition towards fifth generation (5G), mobile cellular networks are evolving into ubiquitous systems for data acquisition and information processing suitable for monitoring and control of large-scale systems. At the forefront of this evolution is the transformation of radio access network (RAN) to support massive-scale machine-type communications (MTC) \cite{3gpp-mtc} and transformation of core network (CN) to support large-scale centralized or distributed information processing through Cloud-RAN (C-RAN) and Fog-RAN (F-RAN) architecture \cite{c-ran}, \cite{f-ran}. MTC services in 5G will offer both massive-scale data acquisition from various machine-type devices through massive MTC (mMTC) service, but also, provide ultra reliable and low-latency communication (URLLC) service for mission-critical applications \cite{mtc-5g}. Complemented with ultra-dense RAN deployment and flexible and virtualized signal processing architecture, novel 5G network services that are particularly suitable for large-scale system monitoring and control of various \emph{smart infrastructures} are emerging \cite{5g-sg}.

In this work, we focus on a generic state estimation problem placed in the context of a future 5G-inspired C-RAN-based cellular network. We consider an underlying large-scale physical system characterized by the state vector $\mathbf{s}$ that contains values of $N$ system state variables. The state variables are observed through the set of $M$ measurements $\mathbf{x}$ of physical quantities collected at the measurement devices spread across the system. This paper considers linear system model in which measured quantities are linear functions of the (sub)set of state variables. Further, we assume measurements are wirelessly communicated across C-RAN-based cellular network. In C-RAN, large number of spatially distributed remote radio heads (RRH) constitutes an ultra-dense RAN infrastructure that receives signals from densely populated MTC devices (e.g., the measurement devices under consideration) \cite{c-ran}. The signal vector $\mathbf{y}$ collected at RRHs is forwarded via backhaul links to a central C-RAN location where it is fed into a collection of base-band units (BBU) for signal detection and recovery. In the standard C-RAN signal detection problem, the goal is to recover the signal $\mathbf{x}$ transmitted by the set of MTC devices from the signal $\mathbf{y}$ received at RRHs and gathered centrally at BBUs \cite{c-ran-cm}\cite{c-ran-bp}. However, in this paper, focusing on widely applicable linear system state estimation, we extend this goal and investigate the problem of recovering the system state $\mathbf{s}$ directly from the signal $\mathbf{y}$ collected across the C-RAN.

The problem we observe represents a concatenation of the two well-studied problems: the linear system state estimation problem (see, e.g., \cite{monticelli}, for the case of power system state estimation) and the problem of uplink signal detection in C-RAN \cite{c-ran-cm}. For the joint problem, it is straightforward to derive (and implement at a central location) the standard minimum mean-square error (MMSE) estimator, however, such a solution comes with prohibitive $O(N^3)$-complexity that hinders its application for large-scale systems. By exploiting inherent sparsity within both of the component problems, an approximate MMSE solution for each problem can be obtained using the tools from probabilistic graphical models, as recently investigated for both (power system) state estimation \cite{se-bp} and uplink signal detection in C-RANs \cite{c-ran-bp}. In particular, an instance of the Belief-Propagation (BP) algorithm, called Gaussian BP (GBP) \cite{gbp}, can be applied to produce an exact MMSE estimate with $O(N)$-complexity, thus scaling the MMSE solution to large-scale system scenarios.

In this paper, we motivate, formulate and solve the linear system state estimation problem considered jointly with the signal detection problem in C-RAN-based cellular networks. We cast the problem of estimating the system state $\mathbf{s}$ from the received vector $\mathbf{y}$ into an equivalent maximum a-posteriori (MAP) problem, and place it into the framework of a popular class of probabilistic graphical model called factor graphs. The state estimate $\hat{\mathbf{s}}$ is then derived as a solution of the GBP algorithm applied over a specific bi-layer structure of the factor graph. Throughout the paper, we use state estimation in power systems with the measurements collected via 5G-inspired C-RAN network as a running example. Our initial numerical results demonstrate the viability of the proposed approach, both in terms of accuracy and convergence.

The paper is organized as follows. In Section II, we present the joint state estimation and C-RAN uplink communication system model. In Section III, this model is mapped into a corresponding factor graph, and the state estimate is obtained via GBP. Section IV provides numerical results of the proposed GBP state estimator. The paper is concluded in Section V. 

\section{System Model}

\begin{figure*}[ht]
	\centering
	\includegraphics[width=17cm]{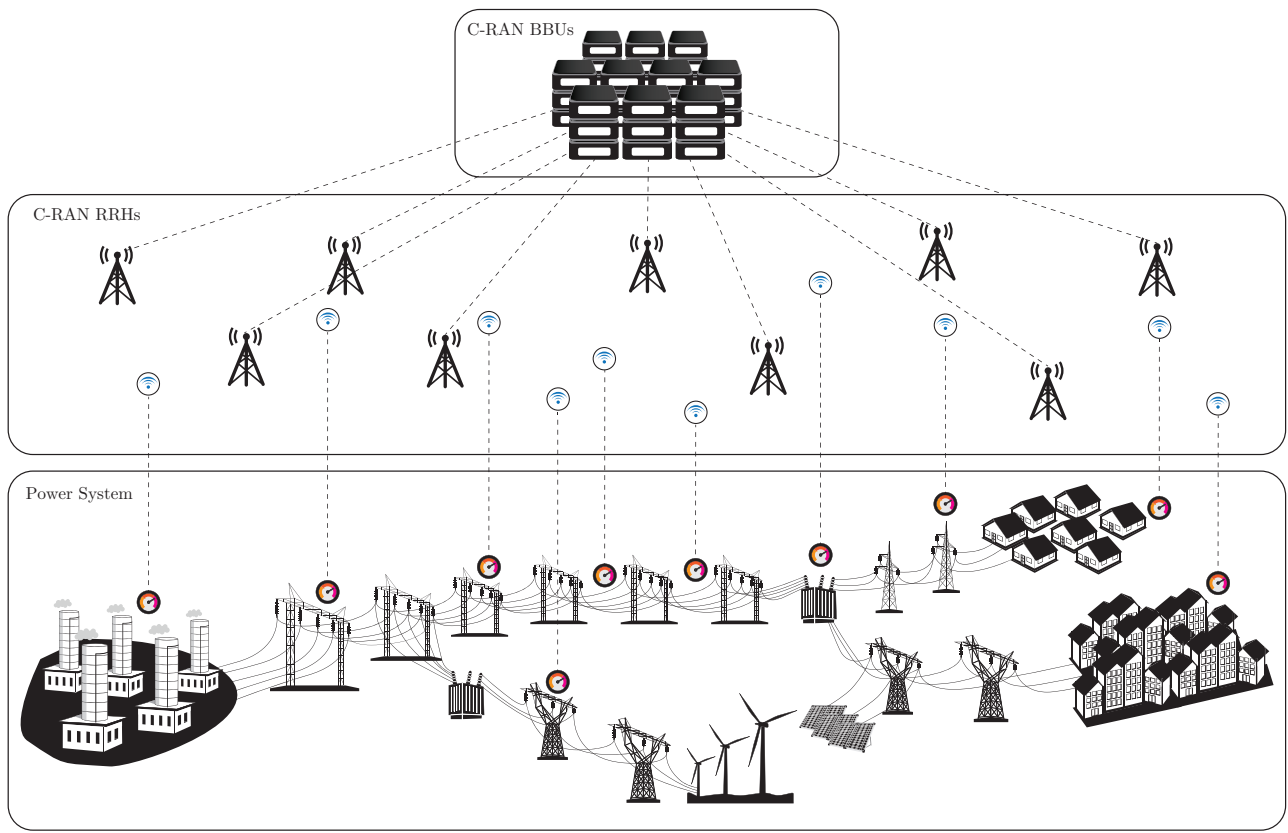}
	\caption{System model example: State estimation in Smart Grid via C-RAN-based mobile cellular network.}
	\label{Figure_1}
	\end{figure*} \noindent

We consider a generic state estimation problem where a set of state variables $\mathbf{s}$ is estimated from a set of observed noisy linear measurements $\mathbf{x}$. However, unlike the traditional setup where the measurements in $\mathbf{x}$ are assumed available at a central node, here we assume they are transmitted via radio access network (RAN) of a mobile cellular system based on a cloud-RAN (C-RAN) architecture. The received signal $\mathbf{y}$ is collected at a large-number of densely deployed remote radio heads (RRHs) and jointly processed at the C-RAN base-band units (BBUs). The problem we consider is that of estimating the system state $\mathbf{s}$ from the received signal $\mathbf{y}$.

\textbf{Linear system measurements model:} We consider a system described via the set of $N$ state variables $\mathbf{s}=(s_1, s_2, \ldots, s_N)^T \in \mathbb{C}^{N \times 1}$. The system is observed via a set of $M$ measurements $\mathbf{x}=(x_1, x_2, \ldots, x_M)^T \in \mathbb{C}^{M \times 1}$. Each measurement is a linear function of the state variables additionally corrupted by the additive noise, i.e., $x_i = \mathbf{a}_i \cdot \mathbf{s} + n_i$, $1 \leq i \leq M$, where $\mathbf{a}_i=(a_{i,1}, a_{i,2}, \ldots, a_{i,N}) \in \mathbb{C}^{1 \times N}$ is a vector of coefficients, while $n_i \in \mathbb{C}$ is a complex random variable. Overall, the system is represented via noisy linear observation model $\mathbf{x}=\mathbf{A}\cdot \mathbf{s} + \mathbf{n}$, where the matrix $\mathbf{A} \in \mathbb{C}^{M \times N}$ contains vectors $\mathbf{a}_i, 1 \leq i \leq M,$ as rows, while $\mathbf{n}=(n_1, n_2, \ldots, n_M)^T \in \mathbb{C}^{M \times 1}$ is a vector of additive noise samples. We assume noise samples $n_i$ are independent identically distributed (i.i.d.) Gaussian random variables with variance $\sigma_{n_i}^2$. For simplicity, we assume the measurement noise variances are equal, i.e., $\sigma_{n_i}^2 = \sigma_{n}^2, 1 \leq i \leq M$. In other words, $\mathbf{n}$ represents a complex Gaussian random vector with zero means $\mathbf{\mu}_{\mathbf{n}} = \mathbf{0}$ and the covariance matrix $\mathbf{\Sigma}_{\mathbf{n}}=\sigma_n^2 \mathbf{I} \in \mathbb{C}^{M \times M}$ ($\mathbf{I}$ is an identity matrix).

\textbf{C-RAN uplink communication model:} In the standard state estimation models, measurements are either assumed available, or they are communicated to the central node, where the state estimation problem is solved. Communication models typically involve point-to-point communication links between the measurement devices and the central node, affected by communication impairments such as delays, packet losses, limited bit rates, etc. In the cellular networks context, this assumes reservation of uplink resources and subsequent non-orthogonal transmission, which typically incurs significant communication delays. Inspired by the recent evolution of massive MTC and ultra-dense C-RAN architectures in upcoming 5G mobile cellular networks, in this work, we consider different grant-less and non-orthogonal communication model, as we detail next. Note that the following C-RAN cellular network model could provide ultra-low latency for the state estimation application under consideration. In other words, such an architecture could produce the system state estimate at the central network node with very low delay after the measurements are acquired, which is crucial for emerging mission-critical 5G MTC use cases \cite{urllc-5g}.  

In the mobile cellular system under consideration, measurements are collected by the measurement devices that we refer to as MTC user equipment (MTC-UE). We consider uplink transmission of $M$ single-antenna MTC UEs towards $L$ single-antenna RRHs. We assume both MTC UEs and RRHs are randomly and uniformly distributed across a given geographic area (note that this placement model is somewhat refined in the numerical results  section). The signal $\mathbf{x} \in \mathbb{C}^{M \times 1}$, representing the set of collected noisy measurements, is transmitted\footnote{Note that, at this point, one can insert specific linear modulation scheme $\mathbf{x}_m = f_{m}(\mathbf{x})$. For simplicity, we assume $f_m(\mathbf{x})=\mathbf{x}$.} by $M$ MTC UEs, while the signal $\mathbf{y}=(y_1, y_2, \ldots, y_L) \in \mathbb{C}^{L \times 1}$ is received at $L$ RRHs, where $\mathbf{y}=\mathbf{H}\cdot \mathbf{x} + \mathbf{m}$. The matrix $\mathbf{H} \in \mathbb{C}^{L \times M}$ represents the channel matrix, where $h_{i,j}$ represents a complex channel coefficient between the $j$-th MTC UE and the $i$-th RRH, while $\mathbf{m}=(m_1, m_2, \ldots, m_{L}) \in \mathbb{C}^L$ is a vector of additive noise samples. As for the measurement process, for the communication process we also assume noise samples $m_i$ are i.i.d. zero-mean Gaussian random variables with variance $\sigma_m^2$, i.e., the mean value and the covariance matrix of $\mathbf{m}$ is given as $\mathbf{\mu}_{\mathbf{m}}=\mathbf{0}$ and $\mathbf{\Sigma}_{\mathbf{m}}=\sigma_m^2 \mathbf{I} \in \mathbb{C}^{L \times L}$, respectively.         

\textbf{Linear system state estimation problem:} From the received signal $\mathbf{y}$ collected across RRHs, we are interested in finding an estimate $\mathbf{\hat{s}}$ of the state vector $\mathbf{s}$. In this paper, we focus on the centralized C-RAN architecture, where all the BBUs are collocated at the central C-RAN node. Thus, due to availability of $\mathbf{y}$ at the central location, we consider centralized algorithms for the state estimation problem. However, the solution we propose in this paper is based on GBP framework, thus it is easily adaptable to a distributed scenario that we refer to as fog-RAN (F-RAN), where BBUs are distributed across different geographic locations closer to the MTC UEs. We refer the interested reader to our recent overview of distributed algorithms for solving the state estimation problem in the context of upcoming 5G cellular networks \cite{5g-sg}.

A common centralized approach to solve the state estimation problem is to provide the minimum mean-square error (MMSE) estimate. One can easily obtain the MMSE estimate $\mathbf{\hat{s}}$ for the underlying linear model in the form:
\begin{equation}\label{MMSE}
\mathbf{\hat{s}}=\big[\mathbf{\Sigma}_{s}^{-1}+ (\mathbf{H}\mathbf{A})^H\mathbf{\Sigma}^{-1} \cdot (\mathbf{H}\mathbf{A})\big]^{-1}
(\mathbf{H}\mathbf{A})^H \mathbf{\Sigma}^{-1} \mathbf{y},
\end{equation} 
where $(\cdot)^H$ is the conjugate-transpose matrix operation, $\mathbf{\Sigma}=\mathbf{H}\mathbf{\Sigma}_{\mathbf{n}}\mathbf{H}^H + \mathbf{\Sigma}_{\mathbf{m}}$, and where we assume prior distribution of $\mathbf{s}$ is i.i.d. Gaussian with mean $\mathbf{\mu}_{\mathbf{s}}=\mathbf{0}$ and $\mathbf{\Sigma}_{\mathbf{s}}=\sigma_s^2 \mathbf{I} \in \mathbb{C}^{N \times N}$. However, solving \eqref{MMSE} scales as $O(N^3)$ which makes linear MMSE state estimation inapplicable in large-scale systems which are of interest in this paper.

In the following, we cast the MMSE estimation into an equivalent MAP state estimator as follows:
\begin{equation}\label{MAP}
\mathbf{\hat{s}}=\arg\max_{\mathbf{s} \in \mathbb{C}^N} P(\mathbf{s}|\mathbf{y}),
\end{equation}
where $P(\mathbf{s}|\mathbf{y})$ is the posterior probability of the state $\mathbf{s}$ after the signal $\mathbf{y}$ is observed at the RRHs. As we will demonstrate in the sequel, if certain sparsity arguments are applicable in the system model under consideration, the solution of the MAP problem can be efficiently calculated using the framework of factor graphs and belief-propagation (BP) algorithms.

\textbf{System model example (Smart Grid):} Before continuing, it is useful to consider an example of the above state estimation setup. We consider the state estimation problem in an electric power system, where the goal is to estimate the state of the power system $\mathbf{s}$, containing complex voltages of $N$ system buses, via the set of measurements $\mathbf{x}$ obtained using measurement devices. Measurement devices are geographically distributed across the power system and we assume they are equipped with wireless cellular interfaces, i.e., they represent MTC UEs connected to the C-RAN based cellular network. The signal $\mathbf{y}$ received from the set of RRHs densely deployed across the cellular network coverage area is processed centrally within the C-RAN system architecture. Fig. \ref{Figure_1} illustrates the smart grid example that we will further refine in Section IV and use as a running example throughout this paper.

\section{State Estimation via Gaussian Belief Propagation}

In this section, we provide a solution to the combined state estimation and uplink signal detection problem defined in \eqref{MAP}, by applying factor graphs and GBP framework. In fact, for both constituent scenarios: the conventional state estimation (in case of power systems) and the uplink signal detection in C-RAN, the GBP has already been proposed and analyzed (see details in \cite{se-bp} and \cite{c-ran-bp}). Thus in this work, we propose using GBP in a joint and combined setup of extracting the system state directly from the observed C-RAN signals.

We note that the various properties of the proposed GBP approach (e.g., complexity, convergence, etc.) will strongly depend on the structure of the underlying factor graph. For example, in terms of complexity (we will come back to convergence in the next section), for GBP to scale well to large-scale systems, it is fundamental that both matrices $\mathbf{A}$ and $\mathbf{H}$ defining the two linear problems are sparse, i.e., that for both $\mathbf{A}$ and $\mathbf{H}$, the number of non-zero entries scales as $O(N)$. In many real-world scenarios, the sparsity typically arises from geographic constraints and reflect \emph{locality} that is typically present in both the measurement and the communication part of the system model. More detailed account on the sparsity of matrices $\mathbf{A}$ and $\mathbf{H}$ clearly depends on the specific scenario under consideration, and we relegate these details to Section IV where we will explicitly deal with the smart grid example introduced earlier. 

\textbf{Factor Graph System Representation:} Assuming that the system state $\mathbf{s}$ has a Gaussian prior, and given that the measurement and communication noise is assumed Gaussian, the MAP problem can be rewritten as follows:
\begin{eqnarray}
\mathbf{\hat{s}} & = &\arg\max_{\mathbf{s} \in \mathbb{C}^N} P(\mathbf{s}|\mathbf{y}) \propto \arg\max_{\mathbf{s} \in \mathbb{C}^N} P(\mathbf{s},\mathbf{y}) = \\ & = & \arg\max_{\mathbf{s} \in \mathbb{C}^N, \mathbf{x} \in \mathbb{C}^M} P(\mathbf{s},\mathbf{x},\mathbf{y}) 
\end{eqnarray}
Note that the distribution $P(\mathbf{s},\mathbf{x},\mathbf{y})$ is jointly Gaussian, and in addition, due to the problem structure where $\mathbf{s}$ and $\mathbf{y}$ are conditionally independent given $\mathbf{x}$, we obtain:
\begin{eqnarray}
\mathbf{\hat{s}} & = & \arg\max_{\mathbf{s} \in \mathbb{C}^N, \mathbf{x} \in \mathbb{C}^M} P(\mathbf{y}|\mathbf{s},\mathbf{x})P(\mathbf{s},\mathbf{x}) = \\
& = & \arg\max_{\mathbf{s} \in \mathbb{C}^N, \mathbf{x} \in \mathbb{C}^M} P(\mathbf{y}|\mathbf{x})P(\mathbf{x}|\mathbf{s})P(\mathbf{s}).
\end{eqnarray}
     
As noted before, in many real-world systems of interest, a measurement $x_j$ is a linear function of a small subset of \emph{local} state variables $\mathbf{s}_{\mathcal{N}(x_j)}$, where $\mathcal{N}(x_j)$ is the index set of the state variables that affect $x_j$, and $\mathbf{s}_{\mathcal{N}(x_j)}=\{s_i|i \in \mathcal{N}(x_j)\}$. In other words, the row-vector $\mathbf{a}_j$ has non-zero components only on a small number of positions indexed by the set $\mathcal{N}(x_j)$, thus making the matrix $\mathbf{A}$ sparse\footnote{More precisely, the number of state variables that affect certain measurement is limited by a constant, independently of the size $N$ of the system.}. Using this fact and the fact that the measurements $x_j$ are mutually independent, we obtain:
\begin{equation}    
P(\mathbf{x}|\mathbf{s}) = \prod_{j=1}^{M} P(x_j|\mathbf{s}_{\mathcal{N}(x_j)}).
\end{equation}

On the other hand, in the C-RAN communication part, although in theory the received signal $y_i$ depends on all the transmitted symbols in $\mathbf{x}$, the channel coefficients between a RRH and a geographically distant MTC UE can be considered negligible, thus leading to matrix $\mathbf{H}$ sparsification \cite{c-ran-bp}. Upon distance-based sparsification proposed in \cite{c-ran-bp}, the received symbol $y_i$ depends only on a small number of symbols $\mathbf{x}_{\mathcal{N}(y_i)}$, where $\mathcal{N}(y_i)$ is the index set of symbols transmitted by the set of MTC UEs in geographic proximity of the $i$-th RRH. Taking the channel sparsification into account, we obtain: 
\begin{equation}    
P(\mathbf{y}|\mathbf{x}) = \prod_{i=1}^{L} P(y_i|\mathbf{x}_{\mathcal{N}(y_i)}).
\end{equation}

Finally, assuming that the state vector is apriori given as a set of i.i.d Gaussian random variables, we obtain the final factorized form  of the initial MAP problem:
\begin{multline}\label{factorized_eq}
\mathbf{\hat{s}} = \arg\max_{\mathbf{s} \in \mathbb{C}^N, \mathbf{x} \in \mathbb{C}^M} \prod_{i=1}^{L} P(y_i|\mathbf{x}_{\mathcal{N}(y_i)}) \cdot \\ \prod_{j=1}^{M} P(x_j|\mathbf{s}_{\mathcal{N}(x_j)}) \cdot \prod_{k=1}^{N} P(s_k).
\end{multline}

\begin{figure}[ht]
	\centering
	\includegraphics[width=7cm]{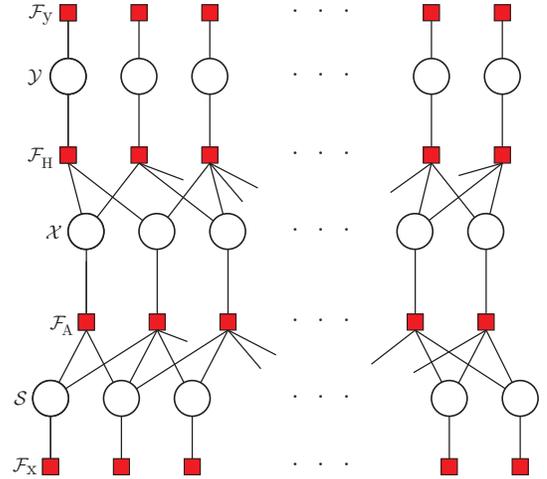}
	\caption{Factor graph representation of the system model.}
	\label{Figure_2}
	\end{figure}

The factor graph representation of the MAP problem follows the factorization presented in \eqref{factorized_eq} and is illustrated in Fig. \ref{Figure_2}. Factor graph $\mathcal{G}=\mathcal{G}(\mathcal{V} \cup \mathcal{F}, \mathcal{E})$ is a bipartite graph consisting of the set of variable nodes $\mathcal{V}$, the set of factor nodes $\mathcal{F}$, and the set of edges $\mathcal{E}$. In our setup, the set $\mathcal{V}$ can be further divided as $\mathcal{V}=\mathcal{S} \cup \mathcal{X} \cup \mathcal{Y},$ where $\mathcal{S}=\{s_1, s_2, \ldots, s_N\}$ is the set of state nodes, $\mathcal{X}=\{x_1, x_2, \ldots, s_M\}$ is the set of measurement nodes, while $\mathcal{Y}=\{y_1, y_2, \ldots, y_L\}$ is the set of received symbol nodes. The set of factor nodes can be divided as $\mathcal{F}=\mathcal{F}_{\mathbf{H}} \cup \mathcal{F}_{\mathbf{A}} \cup \mathcal{F}_{\mathbf{y}}\cup \mathcal{F}_{\mathbf{x}} \cup \mathcal{F}_{\mathbf{s}},$ where $\mathcal{F}_{\mathbf{H}}=\{f_{\mathbf{h}_1}, f_{\mathbf{h}_2}, \ldots, f_{\mathbf{h}_L}\}$ and $\mathcal{F}_{\mathbf{A}}=\{f_{\mathbf{a}_1}, f_{\mathbf{a}_2}, \ldots, f_{\mathbf{a}_M}\}$ represent factor nodes that capture linear relationships between variable nodes described by the rows of matrices $\mathbf{H}$ and $\mathbf{A}$, respectively. In addition, $\mathcal{F}_{\mathbf{y}}=\{f_{y_1}, f_{y_2}, \ldots, f_{y_L}\}$ and $\mathcal{F}_{\mathbf{s}}=\{f_{s_1}, f_{s_2}, \ldots, f_{s_N}\}$ represent the factor nodes that provide inputs due to observations of $\mathbf{y}$ and the prior knowledge about $\mathbf{x}$, respectively, while $\mathcal{F}_{\mathbf{x}}=\{f_{x_1}, f_{x_2}, \ldots, f_{x_M}\}$ serve as virtual inputs needed for initialization of measurement nodes. Similarly, the set of edges $\mathcal{E}$ can be divided as $\mathcal{E}=\mathcal{E}_{\mathbf{H}} \cup \mathcal{E}_{\mathbf{A}} \cup \mathcal{E}_{\mathbf{y}} \cup \mathcal{E}_{\mathbf{x}} \cup \mathcal{E}_{\mathbf{s}}$, where $\mathcal{E}_{\mathbf{H}} \cup \mathcal{E}_{\mathbf{y}} \cup \mathcal{E}_{\mathbf{x}}$ and $\mathcal{E}_{\mathbf{A}} \cup \mathcal{E}_{\mathbf{s}} \cup \mathcal{E}_{\mathbf{x}}$ can be considered as the set of edges of two bipartite subgraphs $\mathcal{G}_{\mathbf{H}}=(\mathcal{Y} \cup \mathcal{X} \cup \mathcal{F}_{\mathbf{H}} \cup \mathcal{F}_{\mathbf{y}} \cup \mathcal{F}_{\mathbf{x}}, \mathcal{E}_{\mathbf{H}} \cup \mathcal{E}_{\mathbf{y}} \cup \mathcal{E}_{\mathbf{x}})$ and $\mathcal{G}_{\mathbf{A}}=(\mathcal{X} \cup \mathcal{S} \cup \mathcal{F}_{\mathbf{A}} \cup \mathcal{F}_{\mathbf{s}} \cup \mathcal{F}_{\mathbf{x}}, \mathcal{E}_{\mathbf{A}} \cup \mathcal{E}_{\mathbf{s}} \cup \mathcal{E}_{\mathbf{x}})$, obtained as the subgraphs of $\mathcal{G}$ induced from the set of factor nodes $\mathcal{F}_{\mathbf{H}} \cup \mathcal{F}_{\mathbf{y}} \cup \mathcal{F}_{\mathbf{x}}$ and $\mathcal{F}_{\mathbf{A}} \cup \mathcal{F}_{\mathbf{s}}\cup \mathcal{F}_{\mathbf{x}}$, respectively\footnote{As noted in footnote 1, if the signal $\mathbf{x}$ is modulated prior to transmission, one can easily add an additional ``layer'' to the factor graph in Fig. \ref{Figure_2} containing a set of $M$ modulated signal variable nodes $\mathcal{X}_m$ connected via modulation factor nodes $\mathcal{F}_m$ with the corresponding measurement nodes $\mathcal{X}$.}. 

As noted earlier, the state estimation problem using GBP over factor graph $\mathcal{G}_{\mathbf{A}}$, and the uplink C-RAN signal detection problem using GBP over factor graph $\mathcal{G}_{\mathbf{H}}$, have been recently investigated in detail in \cite{se-bp} and \cite{c-ran-bp}, respectively. 

\textbf{Gaussian Belief-Propagation and GBP Messages:} To estimate the state variables $\mathbf{s}$, we apply message-passing GBP algorithm \cite{gbp}. GBP operates on the factor graph $\mathcal{G}$ by exchanging messages between factor nodes and variable nodes in both directions. As a general rule, at any variable or factor node, an outgoing message on any edge is obtained as a function of incoming messages from all other edges, using the message calculation rules presented below. In general, the underlying factor graph describing joint state estimation and signal detection problem (Fig. \ref{Figure_2}) will contain cycles, thus the resulting GBP will be iterative, which means that all nodes will iteratively repeat message updates on all of the outgoing edges according to a given message-passing schedule. We provide details on message-passing schedule, correctness and convergence of GBP on loopy graphs later in this section. 

Let us consider a variable node $v_i \in \mathcal{V}$ incident to a factor node $f_j \in \mathcal{F}$. Let $\mathcal{N}(v_i)$ denote the index set of factor nodes incident to $v_i$, and $\mathcal{N}(f_j)$ denote the index set of variable nodes incident to $f_j$. We denote messages from $v_i$ to $f_j$ and from $f_j$ to $v_i$ as $\mu_{v_i \to f_j}(v_i)=(m_{v_i \to f_j},\sigma^2_{v_i \to f_j})$ and $\mu_{f_j \to v_i}(v_i)=(m_{f_j \to v_i},\sigma^2_{f_j \to v_i})$, respectively. Note that, in the GBP scenario, all messages exchanged across the factor graph represent Gaussian distributions defined by the corresponding mean-variance pairs $(m,\sigma^2)$. Thus to describe processing rules in a variable and a factor node, it is sufficient to provide equations that map input $(m,\sigma^2)$-pairs into the output$(m,\sigma^2)$-pair, as detailed below.  
    
\emph{Message from a variable node to a factor node:} the equations below are used to calculate $\mu_{v_i \to f_j}(v_i)=(m_{v_i \to f_j},\sigma^2_{v_i \to f_j})$:
\begin{subequations}
\begin{align}
m_{v_i \to f_j} &= 
\Bigg( \sum_{k \in \mathcal{N}(v_i)\setminus j}
\cfrac{m_{f_k \to v_i}}{\sigma_{f_k \to v_i}^2}\Bigg)\sigma_{v_i \to f_j}^2 \label{BP_vf_mean}\\
\cfrac{1}{\sigma_{v_i \to f_j}^2} &= 		\sum_{k \in \mathcal{N}(v_i)\setminus j} \cfrac{1}{\sigma_{f_k \to v_i}^2}. \label{BP_vf_var}
\end{align}\label{BP_vf_mean_var}	
\end{subequations}

\emph{Message from a factor node to a variable node:} In the setup under consideration, factor nodes represent linear relations between variable nodes. Thus, e.g., for a factor node $f_j$, we can write the corresponding linear relationship as:
\begin{equation}
\begin{gathered}
f_j(\mathbf{v}_{\mathcal{N}(f_j)}) =
C_i v_i + 
\sum_{k \in \mathcal{N}(f_j) \setminus i} 
C_k v_k.
\end{gathered}
\label{BP_general_measurment_fun}
\end{equation}  
With this general notation, the equations below provide $\mu_{f_j \to v_i}(v_i)=(m_{f_j \to v_i},\sigma^2_{f_j \to v_i})$:
\begin{subequations}
\begin{align}
m_{f_j \to v_i} &= 
        \cfrac{1}{C_i} \Bigg(\sum_{k \in \mathcal{N}(f_j) \setminus i} 
        C_k m_{v_k \to f_j}  \Bigg)\label{BP_fv_mean}\\
\sigma_{f_j \to v_i}^2 &=         
        \cfrac{1}{C_i^2} \Bigg(\sum_{k \in \mathcal{N}(f_j) \setminus i} 
        C_k^2 \sigma_{v_k \to f_j}^2  \Bigg). \label{BP_fv_var}
\end{align}\label{BP_fv_mean_var}	
\end{subequations}
		       
\emph{Calculation of marginals:} Applying the above rules in variable and factor nodes of the factor graph results in the sequence of updates of messages exchanged across the edges of the graph. To complete description of loopy GBP, we need to define message initialization at the start, and message scheduling during the course of each iteration, which is done next. After sufficient number of GBP iterations, the final marginal distributions of the random variables corresponding to variable nodes is obtained as: 
\begin{subequations}
\begin{align}
\hat{m}_{v_i} &= 
\Bigg( \sum_{k \in \mathcal{N}(v_i)}
\cfrac{m_{f_k \to v_i}}{\sigma_{f_k \to v_i}^2}\Bigg)\sigma_{v_i}^2 \label{BP_margin_mean}\\
\cfrac{1}{\hat{\sigma}_{v_i}^2} &= 		\sum_{k \in \mathcal{N}(v_i)} \cfrac{1}{\sigma_{f_k \to v_i}^2}. \label{BP_margin_var}
\end{align}\label{BP_margins}	
\end{subequations}

\textbf{GBP Message-Passing Schedule, Correctness and Convergence:} We adopt standard synchronous GBP schedule in which variable node processing is done in the first half-iteration, followed by the factor node processing in the second half-iteration. The iterations are initialized by input messages from $\mathcal{F}_{\mathbf{y}}$ generated from the received signal $\mathbf{y}$, and initial messages from $\mathcal{F}_{\mathbf{x}}$ and $\mathcal{F}_{\mathbf{s}}$ that follow certain prior knowledge (as detailed in the next section).

GBP performance on linear models defined by loopy factor graphs is fairly well understood. For example, if the GBP converges, it is known that the GBP solution will match the solution of the MMSE estimator. The convergence criteria can also be derived in a straightforward manner, by deriving recursive fixed point linear transformations that govern mean value and variance updates through the iterations and investigating spectral radius of such transformations. Due to space restrictions, we leave the details of the convergence analysis in our scenario for the future work.    

\section{Numerical Case Study: Smart Grid State Estimation in 5G C-RAN}
In this section, we specialize our state estimation setup for a case study in which we perform power system state estimation by collecting measurements via 5G-inspired C-RAN. 

\textbf{Power system state estimation - DC model:} For the sake of simplicity, in the following, we consider the linear DC model of a power system. The DC model is an approximate model obtained as a linear approximation of the non-linear AC model that precisely follows the electrical physical laws of the power system. In the DC model, the power system containing $N$ buses is described by $N$ state variables $\mathbf{s}=(s_1, s_2, \ldots, s_N)^T$, where each state variable $s_i=\theta_i$ represents the voltage angle $\theta_i$ (in the DC model, the magnitudes of all voltage phasors are assumed to have unit values). In the DC model, the measurements include only active power flow $P_{rk}$ at the branch $(r,k)$ between the bus $r$ and the bus $k$, active power injection $P_r$ into the bus $r$, and the voltage angle $\theta_r$. Collecting $M$ of such arbitrary measurements across the power system, we obtain the measurement vector $\mathbf{x}=(x_1, x_2, \ldots, x_M)^T$, where each measurement $x_i \in \{P_{rk}, P_r, \theta_r\}$ is a linear function\footnote{More precisely, we have that $P_{rk} = -b_{rk}(\theta_{r}-\theta_{k})$ and $P_{r} = -\sum_{k \in \mathcal{N}_r} b_{rk}(\theta_{r}-\theta_{k})$, where $\mathcal{N}_r$ is the set of adjacent buses of the bus $r$ and $b_{rk}$ is susceptance of the branch $(r,k)$.} of the (sub)set of state variables $\mathbf{s}$, additionally corrupted by additive Gaussian noise of fixed (normalized) noise standard deviation of $\sigma_n$ per unit (p.u.). The noisy measurements $\mathbf{x}$ are then transmitted via C-RAN network as described below. 

\begin{figure}[ht]
	\centering
	\includegraphics[width=8.5cm]{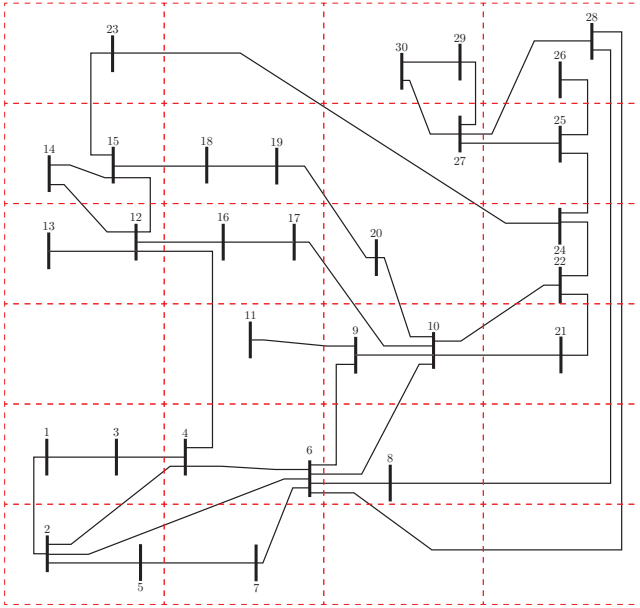}
	\caption{The IEEE 30 bus test case divided into disjoint sub-rectangles.}
	\label{Figure_3}
	\end{figure}

We illustrate the methodology using the IEEE test bus case with 30 buses shown in Fig. \ref{Figure_3} ($N=29$, since one of the bus voltage angles is set to the reference value zero) that we use in the simulations. The example set of $M$ measurements is selected in such a way that the system is observable with the redundancy $M/N$. For each simulation scenario, we generate 1000 random (observable) measurement configurations.  

\textbf{C-RAN cellular network model:}
The set of $M$ MTC-UEs simultaneously transmit their measurements to the set of $L$ RRHs during a given allocated time-frequency slot shared by all MTC-UEs. We assume MTC-UEs and RRHs are placed uniformly at random following independent Poisson Point Process (PPP) in a unit-square area, however, with slight refinement of the PPP placement strategy. Namely, to account for neighboring relations within logical topology of IEEE 30 bus test case, we first divide a unit-square into $w \times q$ disjoint sub-rectangles as shown in Fig. \ref{Figure_3}, and then we assign $M$ MTC-UEs to one of $w\cdot q$ sub-rectangles. We also balance the number of RRHs per sub-rectangle, thus allocating $\sim {L}/(w\cdot q)$ RRHs per sub-rectangle. Finally, all RRHs and MTC-UEs allocated to a given sub-rectangle are placed using the PPP within a given sub-rectangle.

After the placement, we assume $M$ MTC-UEs transmit their signals $\mathbf{x}$, where each measured signal is normalized to its expected normalization value\footnote{We assume normalization constants are known in advance at MTC-UEs and C-RAN nodes, either as a prior knowledge or by long-term averaging.}. For the channel coefficients between the MTC UEs and RRHs, we assume the following model: 
	\begin{equation}    
	h_{i,j} = \gamma_{i,j} d_{i,j}^{-\alpha},
	\end{equation}
where $\gamma_{i,j}$ is the i.i.d. Rayleigh fading coefficient with zero mean and unit variance, $d_{i,j}$ is distance between $i$-th MTC-UE and $j$-th RRH and $\alpha$ is the path loss exponent. We use channel sparsification approach proposed in \cite{c-ran-bp}, with threshold distance set to $d_0={\sqrt{w^2 + q^2}}$ (i.e., equal to the diagonal length of each sub-rectangle). The received signal $\mathbf{y}=(y_1, y_2, \ldots, y_L)$ collected at $L$ RRHs is additionally corrupted by additive Gaussian noise, whose standard deviation is selected so as to provide fixed and pre-defined signal-to-noise ratio (SNR) value. Finally, noisy received signal $\mathbf{y}$ is forwarded via high-throughput backhaul links to C-RAN BBUs. 

\textbf{GBP-based State Estimation:} Using the approach presented in Section III, we apply GBP across the factor graph illustrated in Fig. \ref{Figure_2} to recover the state estimate $\mathbf{x}$ from the received signal $\mathbf{y}$. More precisely, for each random measurement configuration, we generate the part of the factor graph $\mathcal{G}_{\mathbf{A}}$ and, similarly, from known MTC-UE and RRH random positions, we derive\footnote{We note that, in case the small-scale fading is included in the model, one can assume that the channel state information is available at the C-RAN.} the part of the factor graph $\mathcal{G}_{\mathbf{H}}$. Upon reception of $\mathbf{y}$, the GBP runs until it converges. We adopt a synchronous scheduling of GBP messages where messages are synchronously flooded from the factor nodes to variable nodes and back within a single GBP iteration. For a linear model, it is well known that if the GBP converges, it will converge to the minimum mean-square error (MMSE) estimate of the state $\mathbf{x}$.  

\textbf{Simulation Results:} In the first set of experiments, we fix the relative RRH density ${L}/{M}=1$, $\mathrm{SNR}=10$, and redundancy ${M}/{N} = 3$. We investigate the accuracy of the GBP solution of state estimate as a function of different values of measurements noise $\sigma_n=$ $\{10^{-1},$ $10^{-2},$ $10^{-3},$ $10^{-4} \}$. 
\begin{figure}[ht]
	\centering
	\pgfplotstabletranspose[input colnames to=]{\datatable}{./figure/plot1/data.txt}
	\begin{tikzpicture}[]
		\begin{axis}[boxplot/draw direction=y, cycle list name=list2,
   		y tick label style={/pgf/number format/.cd,fixed,
   		fixed zerofill, precision=2, /tikz/.cd},
   		xlabel={Measurement noise $\sigma_n (\mathrm{p.u.})$},
   		ylabel={$\mathrm{RMSE}$},
   		grid=major, 
   		every axis plot/.style={mark=+},  
  		boxplot/every median/.style={red},
   		boxplot/every box/.style={blue},
   		boxplot/every whisker/.style={black},
   		xmin=0, xmax=5, ymin=0, ymax =8,   	
   		xtick={1,2,3,4},
   		xticklabels={$10^{-1}$, $10^{-2}$, $10^{-3}$, $10^{-4}$},
   		ytick={0,2,4,6,8},
   		width=7cm,height=6cm,
   		tick label style={font=\footnotesize}, label style={font=\footnotesize}],
		\pgfplotsinvokeforeach{0,1,2,3}{ 
  		\addplot+[
  		boxplot prepared from table={
    	table=\datatable,
    	row= #1,
    	lower whisker=0,
    	upper whisker=4,
    	lower quartile=1,
    	upper quartile=3,
    	median=2,},
   		boxplot prepared]
  		table [y index=0] {./figure/plot1/outlier#1.txt};}
		\end{axis}
	\end{tikzpicture}
	\caption{The root mean square error of estimate vector of power system state
	 variables obtained by C-RAN and without C-RAN model.}
	\label{plot1}
	\end{figure}
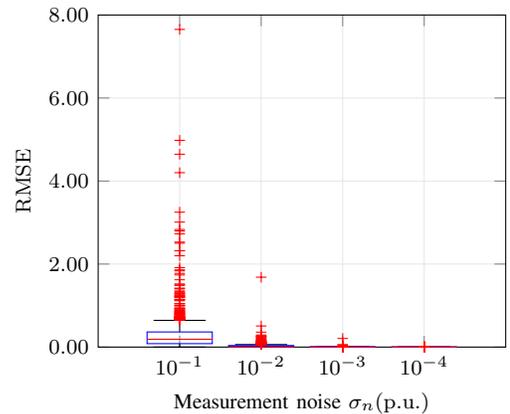

Fig. \ref{plot1} shows the root mean square error $\mathrm{RMSE} = { ({1}/{N}) ||\mathbf {\hat s}_{\mathrm{c}} - \mathbf{\hat s}_{\bar{\mathrm{c}}}||_2 }$, where $\mathbf {\hat s}_{\mathrm{c}}$ and $\mathbf {\hat s}_{\bar{\mathrm{c}}}$ are estimate vectors of power system state variables obtained with and without C-RAN model discussed in this paper, respectively, for different values of measurement noise $\sigma_n$. For the case without C-RAN model, we assume measurements $\mathbf{x}$ are available at BBUs as they are, i.e., without additional noise or errors. In practice, this could be obtained via standard grant-based uplink procedures where each MTC UE is allocated separate orthogonal resources. However, such a strategy incurs significant delay as the underlying system scales, due to message exchange delay, resource allocation delay, as well as ARQ-based error-correction strategies. Note that the C-RAN model described in this paper admits very low latency as all MTC UEs transmit their signals immediately and concurrently.
According to the box plot in Fig. \ref{plot1}, the C-RAN approach is able to reach nearly identical solution as the approach without C-RAN (e.g., $\mathrm{RMSE} \to 0$), if the value of measurements noise is sufficiently low. Note that the typical value (standard deviation) of the measurement noise for devices located across a power system are in the range between $10^{-2}\,\mathrm{ p.u.}$ and $10^{-3}\,\mathrm{ p.u.}$, for legacy measurement devices, and between $10^{-4}\,\mathrm{ p.u.}$ and $10^{-5}\,\mathrm{ p.u.}$, for phasor measurement units. Consequently, the presented approach is suitable for the state estimation in power systems.    
	
In the next simulation experiment, we investigated the system observability as a function of the number of RRHs $L$ deployed in the system, for different values of redundancy $M/N$. We start with $L/M = 0.2$ and increase the RRH density in order to evaluate its effect on the system observability. 
	\begin{figure}[ht]
	\centering
	\begin{tikzpicture}
  	\begin{axis}[width=7cm, height=6cm,
   	x tick label style={/pgf/number format/.cd,
   	set thousands separator={},fixed},
   	legend columns=1,
   	legend style={legend pos=north east,font=\scriptsize, column sep=-0.1cm, 
   	row sep=-0.04cm, legend cell align=left},
   	label style={font=\footnotesize},
   	xlabel={Relative RRH density ${L}/{M}$},
   	ylabel={Unobservable Topologies (\%)},
   	grid=major,
   	xmin = 0.2, xmax = 1.2,
   	xtick={},
   	tick label style={font=\footnotesize}]
	\addlegendimage{legend image with text=$M/N$}
    \addlegendentry{}
	\addlegendentry{1.5} 
	\addlegendentry{2.0}
	\addlegendentry{2.5}
	\addlegendentry{3.0}
	\addlegendentry{3.4}
	
    \addplot[mark=square*,mark repeat=1, mark size=1.5pt, blue] 
   	table [x={ratio}, y={red1_5}] {./figure/plot2/plot2.txt};
   	\addplot[mark=otimes*,mark repeat=1, mark size=1.5pt, red] 
   	table [x={ratio}, y={red2}] {./figure/plot2/plot2.txt};
   	\addplot[mark=triangle*,mark repeat=1, mark size=1.5pt, orange] 
   	table [x={ratio}, y={red2_5}] {./figure/plot2/plot2.txt};   
   	\addplot[mark=diamond*,mark repeat=1, mark size=1.5pt, black] 
   	table [x={ratio}, y={red3}] {./figure/plot2/plot2.txt}; 
   	 \addplot[mark=pentagon*,mark repeat=1, mark size=1.5pt, cyan] 
   	table [x={ratio}, y={redmax}] {./figure/plot2/plot2.txt}; 
  	\end{axis}
 	\end{tikzpicture}
	\caption{The fraction of unobservable system topologies for different values of
	measurement redundancies $M/N$ versus relative RRH density ${L}/{M}$.}
	\label{plot2}
	\end{figure}
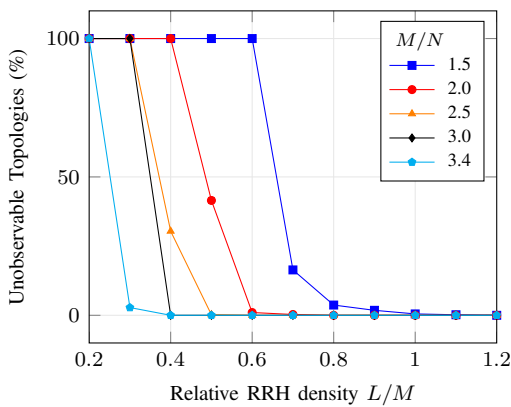

Fig. \ref{plot2} shows the fraction of instances GBP was not able to converge due to insufficient rank of the underlying system as a function of the number of base stations $L$. Note that the fundamental condition for the system to have full rank is that $L \geq N$. By slightly expanding this condition, we get $(L/M) \cdot (M/N) \geq 1$. For all the points in Fig. \ref{plot2} for which this condition is not satisfied, the system is unobservable. If the condition is satisfied, then in each simulation run, a random measurement configuration is verified to provide an observable system, thus the rank insufficiency may only appear as a consequence of the C-RAN topology and the channel matrix sparsification. According to Fig. \ref{plot2}, for the parameters used in our simulations, we can see that GBP generally performs well, however, in the region where $(L/M) \cdot (M/N)$ is slightly above $1$, rank insufficiency may deteriorate the performance. 

Overall, for systems with large number of MTC UEs $M$, the state can be estimated with relatively small number of RRHs $L$. In contrast, for the scenario where a number of MTC UEs is small, the number of RRHs must be increased for successful reconstruction. In addition, simulation results point to capability of the proposed scheme to provide successful reconstruction if the underlying system is observable (i.e., of full rank), while the accuracy of reconstruction (i.e., the accuracy of the state estimator) will depend on the parameters such as channel sparsification, SNR, measurement standard deviations and number of MTC UEs and RRHs. We leave detailed study of these inter-dependencies for our future work.

\section{Conclusions}
Motivated by the development of 5G massive MTC and large-scale distributed 5G C-RAN architecture, in this paper, we proposed a scalable and efficient linear state estimation framework. The proposed framework is based on the GBP algorithm and jointly combines linear state estimation with signal detection in 5G C-RANs. The advantage of GBP solution is accuracy that matches the MMSE estimation, low complexity due to lack of scheduling MTC-UE transmissions, low latency due to simultaneous data transfer, scalability to large-scale systems (due to the fact that the underlying factor graph is usually sparse), and ease of parallelization and distributed implementation in future distributed F-RAN architectures. For the future work, we aim to provide rigorous convergence analysis of GBP in the presented framework, motivated by similar analysis in \cite{c-ran-bp} and \cite{se-bp}, and provide extensive numerical simulation study.

\end{document}